\documentclass[12pt]{iopart}
\usepackage{epsfig}
\begin{document}


\title{On the generalized Hartman effect and transmission time for a particle tunneling through two identical rectangular potential barriers}

\author{N L Chuprikov}

\address{Tomsk State Pedagogical University, 634041, Tomsk, Russia}
\begin{abstract}
We develop a new quantum-mechanical approach to scattering a particle on a one-dimensional (1D) system of two identical rectangular potential
barriers, which implies modelling the dynamics of its subprocesses -- transmission and reflection -- at all stages of scattering. On its basis we
define, for each subprocess, the dwell time as well as the local (exact) and asymptotic (extrapolated) group times. Our concept of the asymptotic
transmission group time confirms the validity of the Wigner phase time in the opaque limit, as well as the existence of the usual and generalized
Hartman effects predicted on its basis. On the energy scale, this concept is valid everywhere in the high energy region as well as in the low
energy region, excepting resonance points and their neighborhoods. On the contrary, the Buttiker dwell time is valid, as the transmission time,
just only at the resonance points. Our concept of the transmission dwell time predicts monotonous growth of the tunneling time when the distance
between the opaque barriers increases. By our approach only this time scale yields the true time spent, on average, by transmitted particles in
the region occupied by the system. We explain why the asymptotic and local transmission group times cannot play this role and why the concept of
{\it transmission group} velocity lies beyond the scope of special relativity. And else, all the transmission times admit only indirect
measurements. Hence the unambiguous interpretation of all tunneling-time experiments is impossible when the transmission dynamics at all stages of
scattering is unknown.
\end{abstract}


\maketitle
\newcommand{\pppp}{\mbox{\hspace{8mm}}}
\newcommand{\ppp}{\mbox{\hspace{5mm}}}
\newcommand{\ooo}{\mbox{\hspace{3mm}}}
\newcommand{\ooa}{\mbox{\hspace{1mm}}}

\section{Introduction} \label{int}

As was shown in \cite{Re1,Re0} for narrow (in $k$ space) wave packets to pass through a 1D system of two identical rectangular potential barriers,
the Wigner phase time does not depend, in the opaque limit, both on the width of the barriers and on the distance between them. This finding,
known in quantum mechanics (QM) and classical electrodynamics (CED) as the generalized Hartman effect, is evident to enforce the tension to appear
due to the usual Hartman effect \cite{Har} between special relativity and the conventional description of the tunneling phenomenon in these two
fundamental theories. Both kinds of the Hartman effect say that either this tunneling time concept to allow superluminal velocities or special
relativity to forbid such velocities must be reconsidered.

The main intrigue is that the Hartman effect is universal (see \cite{Re2,Esp,Re3,Ste1,Ste5}): (a) apart from the Wigner time, it also follows from
the dwell time (see, e.g., \cite{Aha}) and other tunneling-time concepts; (b) it appears not only in QM for non-relativistic and relativistic
particles, but also in classical physics for electromagnetic waves. Besides, the anomalously short tunneling times appear, under some conditions,
even in those approaches which do not predict the saturation of the tunneling time in the opaque limit (see, e.g., the Salecker-Wigner-Peres
timekeeping procedure \cite{Lun}). But, of course, of most importance is the fact that superluminal tunneling velocities are observed
experimentally (see, e.g, \cite{Lon0,Lon,Nim0,Ste,Ste2,Pol,Nim1}). Thus, what has been found in the theoretical approaches \cite{Re1,Re0,Har}
seems to be indeed a real physical effect that only needs the unambiguous interpretation, able to reconcile the observed superluminal tunneling
velocities with special relativity.

One (see, e.g., \cite{Re4}) of the most prominent ideas of solving this problem is to explain superluminal tunneling on the basis of "extended
(non-restricted) special relativity" (see a review \cite{Re6}) with its "switching rule" for tachyon-like particles. A similar point of view was
put forward by Nimtz \cite{Nim} who stated that tunneling is beyond the "jurisdiction" of (the usual, "restricted") special relativity, and
tunnelling takes place due to virtual particles (electrons, photons, etc).

However, some proponents of special relativity prefer another widely spread idea of explaining the observed superluminal group tunneling
velocities (see, e.g., \cite{Re5,Ste1,Kuz}). To diminish the physical significance of such velocities, they put in question the physical
significance of the very concept of the group velocity. As was said in \cite{Kuz}, "causality only requires that the signal velocity of light be
limited by $c$, instead of the group velocity". By the former is meant the velocity of an abrupt leading front of a light pulse, which is always
subluminal (see \cite{Kuz,Re5,Ste2,Aha}). The signal velocity is associated in these approaches with information transfer, and only this velocity
concept is considered to be under the jurisdiction of special relativity.

But again, such a privileged status of the signal velocity was put in doubt by Nimtz and Haibel who stressed in \cite{Nim2} that "A physical
transmitter produces signals of finite spectra only\ldots [Hence f]ront of a signal has no physical meaning\ldots Only the complete envelope\ldots
is the appropriate signal description". As regards observed superluminal tunneling velocities, Nimtz and Haibel show that "The finite duration of
a signal is the reason that a superluminal velocity does not violate the principle of causality". Such a velocity violates special relativity but
is in a full agreement with the non-restricted special relativity \cite{Re6}.

At first glance, these arguments are indeed a sufficient reason for giving up the usual (restricted) special relativity. However, this is not.
Before making a final decision, one has first to ensure that the existing timekeeping procedures, presented in the tunneling time literature,
leave no loophole for the appearance of nonphysical velocities. At the same time, as was shown in \cite{Ch1,Ch6,Ch7} (see also \cite{Ch4,Ch3}),
such a loophole exists.

The point is that the existing quantum-mechanical model of scattering a particle on a 1D potential barrier does not allow tracing the tunneling
dynamics at all stages of scattering. At the initial stage of scattering it shows the incident wave packet that describes the ensemble of
particles to impinge the barrier (from the left, for instance), without distinguishing to-be-transmitted and to-be-reflected particles. At the
very stage of scattering, it shows the process of splitting the incident packet into two parts, again without distinguishing them. And only at the
final stage, this model shows the transmitted and reflected wave packets occupying macroscopically distinct spatial regions (of course, this takes
place only in the case of a one-dimensional {\it completed} scattering (OCS), when the rate of diverging the transmitted and reflected wave
packets exceeds the rate of widening each packet).

As is seen, this model allows one to define the time of arrival of the "center of mass" (CM) of the transmitted wave packet at the right extreme
point of any asymptotically large spatial interval to include the barrier region (by definition, the CM's position is the average value of the
particle's position operator). However, within this model it is impossible to trace its dynamics at the initial stage of scattering and hence to
define the time of departure of this CM from the left extreme point of the interval; the incident wave packet to describe the whole ensemble of
particles has no causal relationship with the transmitted wave packet \cite{La2} (see also \cite{Win}). This fact is well known, but it has not
been taken into account when the concept of the Wigner phase time has been used for studying 1D potential barriers, and namely this incident wave
packet has been used as a counterpart to the transmitted one in the determination of the group-delay time for tunneling. Note that Wigner's paper
\cite{Wig} does not contain this drawback because it deals with the problem of scattering a particle on a point-like scatterer, where there is
only one scattering channel -- reflection.

In our opinion, this step in the phase time concept violates the causality principle and hence opens a loophole for the appearance of superluminal
group tunneling velocities. In order to close it one has to prove that the departure time of the CM of the incident wave packet does coincide with
that of the CM of the wave packet which represents the counterpart to the transmitted one at the initial stage of scattering. To do this, one
needs to restore the whole prehistory of the subensemble of transmitted particles according to its final state. The idea of such reconstruction
has been put forward in \cite{Ch1,Ch6,Ch7} (see also \cite{Ch4,Ch3}) by the example of symmetric potential barriers.

Here we apply this idea to symmetric two-barrier structures and analyze the generalized Hartman effect on its basis. In doing so, we will dwell in
detail on the key points of our approach \cite{Ch1,Ch6,Ch7} in order to make this paper readable on its own.

\section{Backgrounds} \label{back}

Following this approach we begin with the stationary scattering problem. Let a particle with a given momentum $\hbar k$ ($k>0$) be incident from
the left on a system of two identical rectangular potential barriers that occupy the intervals $[a_1,b_1]$ and $[a_2,b_2]$ located to the right of
the origin of coordinates; $0<a_1<b_1<a_2<b_2$. The height of barriers is $V_0$, $b_1-a_1=b_2-a_2=d$ is the width of barriers; $a_2-b_1=L$ is the
distance between them; $b_2-a_1=D$ is the width of the whole two-barrier system. The only difference between this model and \cite{Ch1,Ch6,Ch7} is
that now we deal with the potential function $V(x)$ which is not smooth inside the region $[a_1,b_2]$; the intervals $[a_1,b_1]$, $[b_1,a_2]$ and
$[a_2,b_2]$ should be handled separately.

The wave function $\Psi_{tot}(x,k)$ that describes the stationary state of the ensemble of particles taking part in the process can be written as
follows:
\begin{eqnarray} \label{1}
\fl \Psi_{tot}(x,k)= \left\{
\begin{array}{rl}
e^{ikx}+b_{out}e^{ik(2a_1-x)}:\ppp x\in (-\infty, a_1]\\
a_{tot}^{(1)}\sinh[\kappa(x-a_1)]+b_{tot}^{(1)}\cosh[\kappa(x-a_1)]:\ppp x\in[a_1,b_1]\\
a_{tot}^{gap}\sin[k(x-x_c)]+b_{tot}^{gap}\cos[k(x-x_c)]:\ppp x\in[b_1,a_2]\\
a_{tot}^{(2)}\sinh[\kappa(x-b_2)]+b_{tot}^{(2)}\cosh[\kappa(x-b_2)]:\ppp x\in[a_2,b_2]\\
a_{out}e^{ik(x-D)}:\ppp x\in[b_2,\infty)
\end{array} \right.
\end{eqnarray}
here $\kappa=\sqrt{2m(V_0-E)}/\hbar$; $E=(\hbar k)^2/2m$; $x_c=(b_2+a_1)/2$. We have to stress that the formalism presented is valid not only for
$E<V_0$ (when the Hartman effect emerges) but also for $E\geq V_0$ (in this case, $\kappa$ becomes purely imaginary, with all the consequences).

In order to find the unknown coefficients in Exps. (\ref{1}) we will use the transfer matrix method \cite{Ch2}. By the well known transfer matrix
approach, the expressions $\Psi(x,k)=A_{left}\exp(ikx)+B_{left}\exp(-ikx)$ and $\Psi(x,k)=A_{right}\exp(ikx)+B_{right}\exp(-ikx)$ -- solutions to
the Schr\"{o}dinger equation in the free spaces $x<a$ and $x>b$ for any potential barrier located in the finite interval $[a,b]$ -- are linked by
the transfer matrix,
\begin{eqnarray} \label{2}
\fl \left(
\begin{array}{rl}
A_{left}\\ B_{left}
\end{array} \right) =\textbf{Y}_{(a,b)}
\left(
\begin{array}{rl}
A_{right} \\ B_{right}
\end{array} \right)\ooo
\textbf{Y}_{(a,b)}=\left(
\begin{array}{rl}
q_{(a,b)} & p_{(a,b)} \\ p^*_{(a,b)} & q^*_{(a,b)}
\end{array} \right).
\end{eqnarray}
According to \cite{Ch2}, the elements of the transfer matrix $\textbf{Y}_{(a,b)}$ are determined as follows
\begin{eqnarray*}
\fl q_{(a,b)}=\frac{1}{\sqrt{T_{(a,b)}}}\exp\left\{i\left[k(b-a)-J_{(a,b)}\right]\right\},\ooa p_{(a,b)}=i\sqrt{\frac{R_{(a,b)}}{T_{(a,b)}}} \ooa
\exp\left\{i\left[F_{(a,b)}-k(b+a)\right]\right\}
\end{eqnarray*}
The (real) transmission coefficient $T_{(a,b)}$ and two phases $J_{(a,b)}$ and $F_{(a,b)}$ are determined by explicit analytical expressions when
the barrier is rectangular or $\delta$-potential; when the barrier represents, in its turn, a many-barrier structure, these scattering parameters
obey recurrence relations (see \cite{Ch2}); in all cases $R_{(a,b)}=1-T_{(a,b)}$. For any symmetric structure the phase $F_{(a,b)}$ can take only
two values, either $0$ or $\pi$ (see \cite{Ch2}).

Thus, for the above two-barrier system and its left and right barriers the corresponding transfer matrices $\textbf{Y}_{two}$, $\textbf{Y}_1$ and
$\textbf{Y}_2$
 can be written in the form
\begin{eqnarray} \label{3}
\fl \textbf{Y}_n=\left(
\begin{array}{rl}
q_n & p_n \\ p^*_n & q^*_n
\end{array} \right),\ppp
\textbf{Y}_1\textbf{Y}_2=\textbf{Y}_{two}=\left(
\begin{array}{rl}
q_{two} & p_{two} \\ p^*_{two} & q^*_{two}
\end{array} \right)
\end{eqnarray}
where $q_n=q\cdot\exp[ik(b_n-a_n)]$, $p_n=ip\cdot\exp[-ik(b_n+a_n)]$ $(n=1,2)$;
\begin{eqnarray*}
\fl q=\frac{e^{-iJ}}{\sqrt{T}},\ooo p=\sqrt{\frac{R}{T}}\ooa e^{iF};\ppp q_{two}=\frac{1}{\sqrt{T_{two}}}e^{i[k(b_2-a_1)-J_{two}]},\ooo
p_{two}=i\sqrt{\frac{R_{two}}{T_{two}}}\ooa e^{i[F_{two}-k(b_2+a_1)]}
\end{eqnarray*}
For rectangular barriers the one-barrier parameters $T$, $J$ and $F$ are determined by the expressions (see \cite{Ch2})
\begin{eqnarray*}
\fl T=\left[1+\theta_{(+)}^2 \sinh^2(\kappa d)\right]^{-1},\ooa J=\arctan\left(\theta_{(-)}\tanh(\kappa d)\right)+J^{(0)},\ooa
\theta_{(\pm)}=\frac{1}{2}\left(\frac{k}{\kappa}\pm \frac{\kappa}{k}\right);
\end{eqnarray*}
$J^{(0)}=0$, if $\cosh(\kappa d)>0$; otherwise, $J^{(0)}=\pi$ (this can occur for $E\geq V_0$); $F=0$, if $\theta_{(+)}\sinh(\kappa d)>0$;
otherwise, $F=\pi$. From the latter it follows that $p$ is a real quantity; it can be rewritten as $p=\eta \sqrt{R/T}$; here $\eta=+1$, if
$\theta_{(+)}\sinh(\kappa d)>0$; otherwise, $\eta=-1$.

The two-barrier parameters $T_{two}$, $J_{two}$ and $F_{two}$ are determined by Eq. (\ref{3}) (see the recurrence relations for the scattering
parameters in \cite{Ch2}):
\begin{eqnarray} \label{4}
\fl T_{two}^{-1}=1+4\frac{R}{T^2}\cos^2\chi,\ooo J_{two}=J+\arctan\left(\frac{1-R}{1+R}\tan\chi\right)+F_{two}^{(0)},\ooo F_{two}=F+F_{two}^{(0)};
\end{eqnarray}
here $\chi=J+kL$; $F_{two}^{(0)}=0$, if $\cos\chi\geq 0$; otherwise, $F_{two}^{(0)}=\pi$ (the piecewise constant function
$F_{two}(k)$ is discontinuous at the resonance points where $T_{two}=1$).

Now we can write the searched-for coefficients in (\ref{1}) in terms of one-barrier and two-barrier parameters of scattering. For this purpose it
is suitable to rewrite the wave function $\Psi_{tot}(x,k)$ in the interval $[b_1,a_2]$ as
$\Psi_{tot}(x,k)=A_{tot}^{gap}\exp(ikx)+B_{tot}^{gap}\exp(-ikx)$. Then the following relationships are valid
\begin{eqnarray} \label{5}
\fl \left(
\begin{array}{rl}
A_{tot}^{gap}\\ B_{tot}^{gap}
\end{array} \right) =\textbf{Y}_2
\left(
\begin{array}{rl}
a_{out}e^{-ikD)} \\ 0
\end{array} \right) =\textbf{Y}_1^{-1}
\left(
\begin{array}{rl}
1 \\ b_{out}e^{2ika_1}
\end{array} \right).
\end{eqnarray}
As $a_{tot}^{gap}=i\left(A_{tot}^{gap}e^{ikx_c}-B_{tot}^{gap}e^{-ikx_c}\right)$ and
$b_{tot}^{gap}=A_{tot}^{gap}e^{ikx_c}+B_{tot}^{gap}e^{-ikx_c}$, from the first equality in (\ref{5}) it follows that $a_{tot}^{gap}$ and
$b_{tot}^{gap}$ to enter (\ref{1}) are determined by the expressions
\begin{eqnarray} \label{6}
\fl a_{tot}^{gap}= -a_{out}P^*e^{ika_1},\ppp b_{tot}^{gap}=a_{out}Q^*e^{ika_1};
\end{eqnarray}
here $Q=q^*\exp(ikL/2)+ip\ooa \exp(-ikL/2)$, $P=iq^*\exp(ikL/2)+p\ooa\exp(-ikL/2)$. Then, "sewing" the solutions in adjacent intervals at the
points $x=a_1$ and $x=b_2$, we obtain
\begin{eqnarray} \label{7}
\fl a_{tot}^{(1)}=i (1-b_{out})\frac{k}{\kappa}\ooa e^{ika_1},\ooo b_{tot}^{(1)}=(1+b_{out})\ooa e^{ika_1};\ooo a_{tot}^{(2)}=i
a_{out}\frac{k}{\kappa}\ooa e^{ika_1},\ooo b_{tot}^{(2)}=a_{out}e^{ika_1}.\nonumber
\end{eqnarray}

The amplitudes $a_{out}$ and $b_{out}$ can be obtained either through the one-barrier parameters, with making use of the second equality in
(\ref{5}), or through the two-barrier ones, with making use of the relationship
\begin{eqnarray*}
\fl \left(
\begin{array}{rl}
1 \\ b_{out}e^{2ika_1}
\end{array} \right)=\textbf{Y}_{two}
\left(
\begin{array}{rl}
a_{out}e^{-ikD} \\ 0
\end{array} \right).
\end{eqnarray*}
As a result, we have two equivalent forms for each amplitude,
\begin{eqnarray} \label{8}
\fl a_{out}=\frac{1}{2}\left(\frac{Q}{Q^*}-\frac{P}{P^*}\right)=\sqrt{T_{two}}\ooa e^{iJ_{two}},\nonumber\\
\fl b_{out}=-\frac{1}{2}\left(\frac{Q}{Q^*}+\frac{P}{P^*}\right)=-i\sqrt{R_{two}}\ooa e^{i(J_{two}-F_{two})}.
\end{eqnarray}
Both the forms are useful for the decomposition technique presented in the next section.

\section{Stationary wave functions for transmission and reflection} \label{tref}

According to \cite{Ch1}, for any semitransparent two-barrier system the total wave function $\Psi_{tot}(x,k)$ to describe the whole scattering
process can be uniquely decomposed, for any values of $x$ and $k$, into the sum of two 'subprocess wave functions' $\psi_{tr}(x,k)$ and
$\psi_{ref}(x,k)$ which describe the transmission and reflection subprocesses, respectively. Both obey the following requirements:
\begin{eqnarray} \label{101}
\fl \pppp (a)\ooo \psi_{tr}(x,k)+\psi_{ref}(x,k)=\Psi_{tot}(x,k);
\end{eqnarray}

(b) unlike $\Psi_{tot}(x,k)$, either subprocess wave function must have only one outgoing wave and only one incoming wave; in this case the
transmitted wave in (\ref{1}) serves as the outgoing wave in $\psi_{tr}(x,k)$, the reflected one represents the outgoing wave in
$\psi_{ref}(x,k)$;

(c) the incoming wave of either subprocess wave function must be joined 'causally', at some joining point $x_{join}(k)$, to the corresponding
outgoing wave; the word 'causally' means that each (complex-valued) subprocess wave function must be continuous at this point together with the
corresponding probability flow density (rather than with its first spatial derivative).

Analysis shows that these requirements uniquely determine the amplitudes of incoming waves in $\psi_{tr}(x,k)$ and $\psi_{ref}(x,k)$. And, as
expected, they are such that the probability flow density associated with $\psi_{tr}(x,k)$ coincides with that of $\Psi_{tot}(x,k)$, and
$\psi_{ref}(x,k)$ is a currentless wave function. According to the above three requirements, any zero of this function might be taken as a joining
point $x_{join}(k)$. However, the searched-for joining point $x_{join}(k)$ must also play the role of the extreme right turning-point for
reflected particles. Thus, it must be causally linked to the two-barrier system that reflects these particles. Besides, it must play the role of
the turning point for particles not only with a given $k$ but also for closely spaced values of $k$. Thus, we should impose one more requirement
on the subprocess wave functions:

(d) the point $x_{join}(k)$ must coincide with such a zero of the currentless wave function $\psi_{ref}(x,k)$ whose position on the $OX$ axis
depends most weakly on the parameter $k$.

Note that for any {\it symmetric} two-barrier system, one of zeros of the wave function $\psi_{ref}(x,k)$, that obeys the requirements (a)-(c),
coincides with the midpoint of the system for any value of $k$. Since this zero does not at all depend on $k$, for such systems $x_{join}(k)=x_c$.
So that, if $x\geq x_c$, then $\psi_{ref}(x,k)\equiv 0$ and $\psi_{tr}(x,k)\equiv \Psi_{tot}(x,k)$ -- particles, reflected by the symmetric two
barrier system, exist only in the region $x<x_c$.

Note, the fact that each subprocess wave function consists of two different solutions of the Schr\"{o}dinger equation, causally connected at the
midpoint $x_c$, has the following physical justification. From the point of view of classical physics the midpoint of the barrier region of any
symmetric potential barrier is an extreme turning point for particles reflected by the barrier, irrespective of its spatial size and the
particle's mass.

In order to fulfill the correspondence principle, our quantum-mechanical model of the scattering process extends this requirement onto atomic
scales. For this purpose it treats the spatial regions $x<x_c$ and $x>x_c$ as those with different physical contexts: the region $x>x_c$ is
inaccessible for quantum reflected particles impinging the barrier from the left, like for classical ones. In these two regions, quantum particles
taking part in the transmission subprocess move under {\it different physical contexts} and hence constitute {\it different quantum ensembles}.
Thus, on the one hand, the same set of particles is described by the different solutions of the Schr\"{o}dinger equation in the regions $x<x_c$
and $x>x_c$, because different contexts imply different solutions; on the other hand, since these ensembles are associated with the same set of
particles at the different stages of scattering, these solutions must be causally connected at the boundary of these regions.

Calculations yield that in the region $x<x_c$ the wave function $\psi_{ref}(x,k)$ can be written as follows,
\begin{eqnarray} \label{102}
\fl \psi_{ref}(x,k)= \left\{
\begin{array}{rl}
A^{in}_{ref}e^{ikx}+b_{out}e^{ik(2a_1-x)}:\ppp x\in (-\infty, a_1]\\
a_{ref}^{(1)}\sinh[\kappa(x-b_1)]+b_{ref}^{(1)}\cosh[\kappa(x-b_1)]:\ppp x\in[a_1,b_1]\\
a_{ref}^{gap}\sin[k(x-x_c)]:\ppp x\in[b_1,x_c]
\end{array} \right.
\end{eqnarray}
Again, as in the previous section, in order to find the amplitudes to enter these expressions it is suitable to rewrite the function
$\psi_{ref}(x,k)$ in the interval $[b_1,x_c]$ in the form $\psi_{ref}(x,k)=A_{ref}^{gap}\exp(ikx)+B_{ref}^{gap}\exp(-ikx)$. The coefficients in
this expression are determined as
\begin{eqnarray} \label{103}
\fl \left(
\begin{array}{rl}
A_{ref}^{gap}\\ B_{ref}^{gap}
\end{array} \right) =\textbf{Y}_1^{-1}
\left(
\begin{array}{rl}
A^{in}_{ref} \\ b_{out}e^{2ika_1}
\end{array} \right).
\end{eqnarray}
Then, making use of the relationships
\begin{eqnarray} \label{104}
\fl a_{ref}^{gap}=i\left(A_{ref}^{gap}\ooa e^{ikx_c}-B_{ref}^{gap}\ooa e^{-ikx_c}\right),\ooo A_{ref}^{gap}\ooa e^{ikx_c}+B_{ref}^{gap}\ooa
e^{-ikx_c}=0
\end{eqnarray}
we can find the unknown coefficients to enter Exps. (\ref{102}).

From the second equality in (\ref{104}) it follows that $A^{in}_{ref}=-b_{out}Q^*/Q$. Or, taking into account Exps. (\ref{7}), we obtain
\begin{eqnarray} \label{105}
\fl A^{in}_{ref}=b_{out}(b_{out}^*-a_{out}^*)=\sqrt{R_{two}}\left(\sqrt{R_{two}}+i\eta_{two}\sqrt{T_{two}}\right)\equiv
\sqrt{R_{two}}\exp(i\lambda)
\end{eqnarray}
where $\eta_{two}=+1$, if $F_{two}=0$; otherwise, $\eta_{two}=-1$. This means that the phases of the incident waves in $\Psi_{tot}(x,k)$ and
$\psi_{ref}(x,k)$ differ from each other by an amount of $\lambda=\eta_{two}\cdot \arctan\sqrt{T_{two}(k)/R_{two}(k)}$.

Then, taking into account, in (\ref{104}), Exps. (\ref{103}) and (\ref{105}), we obtain
\begin{eqnarray*}
\fl a_{ref}^{gap}=-2P b_{out} a^*_{out}e^{ika_1}.
\end{eqnarray*}
And lastly, by making use of the continuity conditions at the point $x=b_1$, we obtain
\begin{eqnarray*}
\fl a_{ref}^{(1)}=\frac{k}{\kappa}a_{ref}^{gap}\cos\left(\frac{kL}{2}\right),\ppp b_{ref}^{(1)}=-a_{ref}^{gap}\sin\left(\frac{kL}{2}\right).
\end{eqnarray*}

Now, when $\psi_{ref}(x,k)$ has been presented, we can write $\psi_{tr}(x,k)$ as follows: $\psi_{tr}(x,k)=\Psi_{tot}(x,k)-\psi_{ref}(x,k)$. In
particular,
\begin{eqnarray} \label{106}
\fl A^{in}_{tr}=1-A^{in}_{ref}=\sqrt{T_{two}}\left(\sqrt{T_{two}}-i\eta_{two}\sqrt{R_{two}}\right)=
\sqrt{T_{two}}\exp\left[i\left(\lambda-\eta_{two}\frac{\pi}{2}\right)\right].
\end{eqnarray}
As is seen, not only $A^{in}_{tr}(k)+A^{in}_{ref}(k)=1$, but also $|A^{in}_{tr}(k)|^2+|A^{in}_{ref}(k)|^2=1$. It should be
stressed also that
\begin{eqnarray} \label{108}
\fl |\psi_{tr}(x_c-x,k)|=|\psi_{tr}(x-x_c,k)|.
\end{eqnarray}

\section{Time-dependent wave functions for transmission and reflection} \label{alt}

Let us now proceed to the time-dependent process described by the wave packet
\begin{eqnarray} \label{200}
\fl \Psi_{tot}(x,t)=\frac{1}{\sqrt{2\pi}}\int_{-\infty}^\infty A(k)\Psi_{tot}(x,k)e^{-iE(k)t/\hbar}dk
\end{eqnarray}
where $A(k)$ is determined by an initial condition. Here we assume $A(k)$ to be the (real) Gaussian function $A(k)=(2l_0^2/\pi)^{1/4}
\exp\left[-l_0^2(k-\bar{k})^2\right]$. In this case
\begin{eqnarray} \label{201}
\fl \bar{x}_{tot}(0)=0,\ppp \bar{p}_{tot}(0)=\hbar\bar{k},\ppp\overline{x^2}_{tot}(0)=l_0^2;
\end{eqnarray}
hereinafter, for any observable $F$ and time-dependent localized state $\Psi^A_B$ \[\fl
\bar{F}^A_B(t)=\frac{<\Psi^A_B|\hat{F}|\Psi^A_B>}{<\Psi^A_B|\Psi^A_B>}\] (if $\bar{F}^A_B(t)$ is constant its argument will be omitted). We assume
that the parameters $l_0$ and $\bar{k}$ obey the conditions for the OCS, mentioned in Section \ref{int}; i.e., we assume that the rate of
diverging the transmitted and reflected wave packets exceeds the rate of widening each packet. We also assume that the origin of coordinates,
which is the starting point of the CM of the wave packet $\Psi_{tot}(x,t)$, lies far enough from the left boundary of the two-barrier system:
$a_1\gg l_0$.

Besides, let the expression
\begin{eqnarray}  \label{202}
\fl \psi_{tr,ref}(x,t)=\frac{1}{\sqrt{2\pi}}\int_{-\infty}^\infty A(k)\psi_{tr,ref}(x,k)e^{-iE(k)t/\hbar}dk
\end{eqnarray}
give the wave functions $\psi_{tr}(x,t)$ and $\psi_{ref}(x,t)$ to describe, respectively, the time-dependent transmission and reflection
subprocesses. It is evident (see Eq.~(\ref{101})) that the sum of these two functions yields, at any value of $t$, the total wave function
$\Psi_{tot}(x,t)$,
\begin{eqnarray} \label{203}
\fl \Psi_{tot}(x,t)=\psi_{tr}(x,t)+\psi_{ref}(x,t).
\end{eqnarray}

So, at the first stage, the OCS is described by the incident packet
\begin{eqnarray*}
\fl \Psi_{tot}(x,t)\simeq\Psi_{tot}^{inc}(x,t)= \frac{1}{\sqrt{2\pi}}\int_{-\infty}^\infty A(k)\exp[i(kx-E(k)t/\hbar)]dk,
\end{eqnarray*}
and its transmission and reflection subprocesses are described by the wave packets
\begin{eqnarray*}
\fl \psi_{tr,ref}\simeq\psi_{tr,ref}^{inc}= \frac{1}{\sqrt{2\pi}}\int_{-\infty}^\infty A(k)A^{in}_{tr,ref}(k)\exp[i(kx-E(k)t/\hbar)]dk.
\end{eqnarray*}
Considering Exps. (\ref{105}) and (\ref{106}) for the amplitudes of the incident waves in $\psi_{tr}(x,k)$ and $\psi_{ref}(x,k)$, it is easy to
show that
\begin{eqnarray} \label{204}
\fl {\bar{x}}_{tr}^{inc}(0)=-{\overline{\lambda^\prime(k)}}_{tr}^{inc}\equiv -\frac{\int_{-\infty}^\infty \lambda^\prime(k) T_{two}(k) A^2(k)
dk}{\int_{-\infty}^\infty T_{two}(k) A^2(k) dk},\\ \fl {\bar{x}}_{ref}^{inc}(0)=-{\overline{\lambda^\prime(k)}}_{ref}^{inc}\equiv
-\frac{\int_{-\infty}^\infty \lambda^\prime(k)R_{two}(k)A^2(k) dk}{\int_{-\infty}^\infty R_{two}(k)A^2(k) dk};\nonumber
\end{eqnarray}
the prime denotes the derivative on $k$. That is, in the general case the CMs of the wave packets $\Psi_{tot}(x,t)$, $\psi_{tr}(x,t)$ and
$\psi_{ref}(x,t)$ start at $t=0$ from the different spatial points!

Similarly, for the final stage of scattering
\begin{eqnarray*}
\fl \psi_{tr}\simeq\psi_{tr}^{out}= \frac{1}{\sqrt{2\pi}}\int_{-\infty}^\infty A(k)a_{out}(k)e^{i[k(x-D)-E(k)t/\hbar]}dk,\\
\fl \psi_{ref}\simeq\psi_{ref}^{out}= \frac{1}{\sqrt{2\pi}}\int_{-\infty}^\infty A(k)b_{out}(k)e^{i[k(2a_1-x)-E(k)t/\hbar]}dk.
\end{eqnarray*}
Thus, since $|A^{in}_{tr}(k)|^2=|a_{out}(k)|^2=T_{two}(k)$ and $|A^{in}_{ref}(k)|^2=|b_{out}(k)|^2=R_{two}(k)$ (see (\ref{8}), (\ref{105}) and
(\ref{106})), for the initial and final stages of scattering we have
\begin{eqnarray*}
\fl \langle\psi_{tr}^{inc}|\psi_{tr}^{inc}\rangle=\langle\psi_{tr}^{out}|\psi_{tr}^{out}\rangle=\int_{-\infty}^\infty T_{two}(k)A^2(k)dk \equiv
{\textbf{T}}_{as},\\ \fl \langle\psi_{ref}^{inc}|\psi_{ref}^{inc}\rangle=\langle\psi_{ref}^{out}|\psi_{ref}^{out}\rangle =\int_{-\infty}^\infty
R_{two}(k)A^2(k)dk\equiv {\textbf{R}}_{as}.
\end{eqnarray*}

In its turn, since $T_{two}(k)+R_{two}(k)=1$ and $\langle\Psi_{tot}|\Psi_{tot}\rangle=\int_{-\infty}^\infty A^2(k)dk=1$, from the above it follows
that the constant norms ${\textbf{T}}_{as}$ and ${\textbf{R}}_{as}$ give unit in sum:
\begin{eqnarray} \label{205}
\fl {\textbf{T}}_{as}+{\textbf{R}}_{as}=1.
\end{eqnarray}
The fact that at both these stages of scattering the transmission and reflection subprocesses obey the probabilistic "either-or" rule (\ref{205})
means that they behave at these stages as {\it alternative} subprocesses, despite interference to exist between them at the initial stage. This
interference is such that
\begin{eqnarray*}
\fl \langle\psi_{tr}^{inc}|\psi_{ref}^{inc}\rangle=\int_{-\infty}^\infty A^2(k)\left[A^{in}_{tr}(k)\right]^*A^{in}_{ref}(k)dk =
i\int_{-\infty}^\infty A^2(k) \eta_{two}(k)\sqrt{T_{two}(k)R_{two}(k)}dk
\end{eqnarray*}
(the real-valued function $\eta_{two}(k)$ is defined in (\ref{105})). Thus,
$\langle\psi_{tr}^{inc}|\psi_{ref}^{inc}\rangle+\langle\psi_{ref}^{inc}|\psi_{tr}^{inc}\rangle=0$.

At the very stage of scattering, when the wave packet $\psi_{tr}(x,t)$ crosses the point $x_c$, the norm
${\textbf{T}}=\langle\psi_{tr}|\psi_{tr}\rangle$ varies. Fact is that the requirements (a)-(d) (see Section \ref{tref}) ensure the balance of the
input $I_{tr}(x_c-0,k)$ and output $I_{tr}(x_c+0,k)$ probability flows only for each single wave $\psi_{tr}(x,k)$ entering the wave packet
$\psi_{tr}(x,t)$. For the packet itself, these requirements (according to which the first derivative of the wave function $\psi_{tr}(x,k)$ remains
discontinuous) do not ensure the balance of the corresponding (time-dependent) probability flows.

Now $d\textbf{T}/dt=I_{tr}(x_c+0,t)-I_{tr}(x_c-0,t)\neq 0$. This effect takes place due to the nonlinearity of the continuity equation for wave
functions, or, more precisely, due to the interaction of the main 'harmonic' $\psi_{tr}(x,\bar{k})$ with the 'subharmonics' $\psi_{tr}(x,k)$ to
constitute the wave packet $\psi_{tr}(x,t)$. Thus, since the role of subharmonics is essential at the leading and trailing fronts of the
wave-packet, this effect is maximal when one of these fronts crosses the midpoint $x_c$. Of course, the {\it total} variation of the norm
$\textbf{T}$, gained in the course of the whole OCS, is zero.

As regards $\textbf{R}$, this norm remains constant even at the very stage of scattering: $\textbf{R}\equiv\textbf{R}_{as}$. This follows from the
fact that $I_{ref}(x_c+0,t)=I_{ref}(x_c-0,t)=0$ since $\psi_{ref}(x_c,t)=0$ for any value of $t$.

Now, when the transmission and reflection dynamics at all stages of scattering has been revealed, we can proceed to the study of the temporal
aspects of each subprocess. As it will be seen from the following, the unusual properties of the transmission subprocess play crucial role in the
interpretation of the Hartman effect.

\section{The local and asymptotic group scattering times} \label{as}

We begin with the presentation of local (exact) and asymptotic (extrapolated) group times for transmission and reflection. For example, the local
transmission group time $\tau_{tr}^{loc}$ to characterize the dynamics of the CM of the wave packet $\psi_{tr}(x,t)$ in the region $[a_1,b_2]$
occupied by the two-barrier system is defined as follows (see \cite{Ch1}): $\tau_{tr}^{loc}=t_{tr}^{exit}-t_{tr}^{entry}$, where $t_{tr}^{entry}$
and $t_{tr}^{exit}$ are such instants of time that
\begin{eqnarray*}
\fl \bar{x}_{tr}(t_{tr}^{entry}) =a_1,\ppp \bar{x}_{tr}(t_{tr}^{exit})=b_2.
\end{eqnarray*}
Similarly, for reflection $\tau_{ref}^{loc}=t_{ref}^{exit}-t_{ref}^{entry}$, where $t_{ref}^{entry}$ and $t_{ref}^{exit}$ are two different roots,
if any, of the same equation ($t_{ref}^{entry}<t_{ref}^{exit}$):
\begin{eqnarray*}
\fl \bar{x}_{ref}(t_{ref}^{entry}) =a_1,\ppp \bar{x}_{ref}(t_{ref}^{exit})=a_1.
\end{eqnarray*}
If this equation has no more than one root, $\tau_{ref}^{loc}=0$.

The main feature of $\tau_{tr}^{loc}$ and $\tau_{ref}^{loc}$ is that, even for rectangular barriers, these characteristic times can be calculated
only numerically. Moreover, they do not give a complete description of the temporal aspects of each subprocess, because the two-barrier system
affects the subensembles of transmitted and reflected particles not only when the CMs of the wave packets $\psi_{tr}(x,t)$ and $\psi_{ref}(x,t)$
move in the region $[a_1,b_2]$. Of importance is also to define the {\it asymptotic} group times to describe these subprocesses in the
asymptotically large spatial region $[0,b_2+\Delta X]$ where $\Delta X\gg l_0$.

In doing so, we have to take into account that either wave packet does not interact with the system when its CM is at the extreme points of this
region. That is, the {\it asymptotic} transmission time can be defined in terms of the transmitted $\psi_{tr}^{out}$ and to-be-transmitted
$\psi_{tr}^{inc}$ wave packets. Similarly, the asymptotic reflection time can be introduced in terms of the wave packets $\psi_{ref}^{out}$ and
$\psi_{ref}^{inc}$.

We begin with the transmission subprocess. For the CM's position $\bar{x}_{tr}(t)$ at the initial stage of scattering we have (see also
(\ref{204}))
\begin{eqnarray}  \label{301}
\fl \bar{x}_{tr}(t)\simeq{\bar{x}}_{tr}^{inc}(t)=\frac{\hbar {\bar{k}}_{tr}}{m}t-{\overline{\lambda^\prime(k)}}_{tr}^{inc};
\end{eqnarray}
here ${\bar{k}}_{tr}={\bar{k}}_{tr}^{out}={\bar{k}}_{tr}^{inc}$. At the final stage
\begin{eqnarray*}
\fl \bar{x}_{tr}(t)\simeq{\bar{x}}_{tr}^{out}(t)=\frac{\hbar {\bar{k}}_{tr}}{m}t-{\overline{J_{two}^\prime(k)}}_{tr}^{out} +D.
\end{eqnarray*}
Thus, the time $\tau^{gr}_{tr}(0,b_2+\Delta X)$ spent by the CM of $\psi_{tr}(x,t)$ in $[0,b_2+\Delta X]$ is
\begin{eqnarray*}
\fl \tau^{gr}_{tr}(0,b_2+\Delta X)\equiv t_{arr}-t_{dep}=\frac{m}{\hbar \bar{k}_{tr}}\left[{\overline{J_{two}^\prime(k)}}_{tr}^{out}
-{\overline{\lambda^\prime(k)}}_{tr}^{inc}+a_1+\Delta X \right],
\end{eqnarray*}
where the arrival time $t_{arr}$ and the departure time $t_{dep}$ obey the equations
\[\fl {\bar{x}}_{tr}^{inc}(t_{dep})=0;\ooo {\bar{x}}_{tr}^{out}(t_{arr})=b_2+\Delta X.
\]
The quantity $\tau_{tr}^{as}=\tau^{gr}_{tr}(a_1,b_2)$ -- the input of the region $[a_1,b_2]$ -- will be referred to as the asymptotic
(extrapolated) transmission group time:
\begin{eqnarray} \label{302}
\fl \tau^{as}_{tr}=\frac{m}{\hbar \bar{k}_{tr}}\left[{\overline{J_{two}^\prime(k)}}_{tr}^{out}
-{\overline{\lambda^\prime(k)}}_{tr}^{inc}\right].
\end{eqnarray}
Similarly, for reflection we have
\begin{eqnarray} \label{303}
\fl \tau^{as}_{ref}=\frac{m}{\hbar \bar{k}_{ref}}\left[{\overline{J_{two}^\prime(k)}}_{ref}^{out}
-{\overline{\lambda^\prime(k)}}_{ref}^{inc}\right];
\end{eqnarray}
${\bar{k}}_{ref}^{inc}=-{\bar{k}}_{ref}^{out}={\bar{k}}_{ref}$.

For narrow (in $k$-space) wave packets (the value of $l_0$ is large enough)
\begin{eqnarray*}
\fl \tau^{as}_{tr}(k)=\tau^{as}_{ref}(k)\equiv\tau_{as}(k)=\frac{m}{\hbar k}\left[J_{two}^\prime(k)-\lambda^\prime(k)\right];\\
\fl {\bar{x}}_{tr}^{inc}(0)={\bar{x}}_{ref}^{inc}(0)\equiv x_{start}=-\lambda^\prime(k)
\end{eqnarray*}
(the upper line in the notation $\bar{k}$ was omitted)

Note that the above expressions are valid for any symmetric two-barrier system. But only in particular cases, including the case with rectangular
barriers, we can obtain explicit expressions for the above time scales. For the case under study we have
\begin{eqnarray*}
\fl J_{two}^{\prime}=J^\prime+\frac{T_{two}}{T^2}\left[T(1+R)\left(J^\prime+L\right)+
T^\prime\sin[2(J+kL)]\right],\\
\fl \lambda_{two}^{\prime}=2\eta \frac{T_{two}}{\sqrt{R}\ooa
T^2}\left[T^\prime(1+R)\cos(J+kL)+2RT(J^\prime+L)\sin(J+kL)\right];\\ \fl
J^\prime=\frac{T}{\kappa}\left[\theta^2_{(+)}\sinh(2\kappa d)+\theta_{(-)}\kappa d\right],\ooo
T^\prime=2\theta^2_{(+)}\frac{T^2}{\kappa}\left[2\theta_{(-)}\sinh^2(\kappa d) +\kappa d \sinh(2\kappa d)\right].
\end{eqnarray*}

Explicit expressions for $\tau^{as}$ and $x_{start}$ are very cumbersome in the general case. However, for $L=0$, when the two-barrier system is
reduced to a single rectangular barrier of width $D$, we have (see \cite{Ch1})
\[\fl \tau_{as}(k)=\frac{4m}{\hbar k\kappa}\ooa\frac{\left[k^2+\kappa_0^2\sinh^2\left(\kappa D/2\right)\right]
\left[\kappa_0^2\sinh(\kappa D)-k^2 \kappa D\right]} {4k^2\kappa^2+ \kappa_0^4\sinh^2(\kappa D)};
\]
\begin{eqnarray} \label{304}
\fl x_{start}(k)= -2\frac{\kappa_0^2}{\kappa}\ooa \frac{(\kappa^2-k^2)\sinh(\kappa D)+k^2 \kappa D \cosh(\kappa D)} {4k^2\kappa^2+
\kappa_0^4\sinh^2(\kappa D)}.
\end{eqnarray}
where $\kappa_0=\sqrt{2mV_0}/\hbar$ (note, focusing on the Hartman effect we assumed that $V_0>0$; however, the formalism presented is valid also
for $V_0<0$ when both $\kappa_0$ and $\kappa$ are purely imagine quantities).

The key difference between the Wigner phase time $\tau_{ph}(k)=mJ_{two}^\prime(k)/\hbar k$ and $\tau_{as}(k)$ is as follows. The former is based
on the unproven assumption that the incident wave packet $\Psi_{tot}^{inc}(x,t)$, multiplied by the factor ${\textbf{T}}_{as}$, can be treated as
a counterpart of the transmitted wave packet at the initial stage of scattering. In fact, the Wigner time concept implies that the average time
$t_{tr}^{dep}$ of departure of transmitted particles from the point $x=0$ coincides with that of all scattering particles. In the considered
setting of the problem, this assumption means that $t_{tr}^{dep}(k)=t_{tot}^{dep}(k)=0$, resulting in the asymptotic transmission group time
$\tau_{ph}=mJ_{two}^\prime(k)/\hbar k$. But the concept $\tau_{as}(k)$ implies that $t_{tr}^{dep}(k)=m\lambda^\prime(k)/\hbar k$. As a result the
asymptotic transmission group time is defined in our approach by the expression
\[\fl \tau_{as}(k)=\tau_{ph}(k)-t_{tr}^{dep}(k);\] note that $t_{ref}^{dep}(k)=t_{tr}^{dep}(k)\equiv \tau_{dep}$.

However, in the opaque-barrier limit ($E<V_0$, $d\to\infty$ and the value of $E$ is far enough from the points of resonance) this assumption is
quite justified. In this limit $\lambda'(k)\to 0$ and, as a consequence, the time scales $\tau_{ph}$ and $\tau_{as}$ coincide with each other.
This means that the Hartman effect predicted in the existing approaches \cite{Re1,Re0,Har} on the basis of the concept of the Wigner phase time
appears also in our approach.

\section{The dwell times for transmission and reflection} \label{loc}

Our next step is to introduce the dwell times for both subprocesses in the case of the stationary scattering problem. For the two-barrier system
the dwell times $\tau^{dwell}_{tr}$ and $\tau^{dwell}_{ref}$ for transmission and reflection, respectively, are defined as follows
\begin{eqnarray*}
\fl \tau^{dwell}_{tr}= \frac{m}{\hbar k T_{two}}\int_{a_1}^{b_2}\left|\psi_{tr}(x,k)\right|^2dx\equiv
\tau^{(1)}_{tr}+\tau^{gap}_{tr}+\tau^{(2)}_{tr},\\
\fl \tau^{dwell}_{ref}=\frac{m}{\hbar k R_{two}}\int_{a_1}^{x_c} \left|\psi_{ref}(x,k)\right|^2dx\equiv \tau^{(1)}_{ref}+\tau^{gap}_{ref};
\end{eqnarray*}
here $\tau^{(1)}_{tr}$ and $\tau^{(1)}_{ref}$ describe the left rectangular barrier located in the interval $[a_1,b_1]$; $\tau^{gap}_{tr}$ and
$\tau^{gap}_{ref}$ characterize the free space $[b_1,a_2]$; $\tau^{(2)}_{tr}$ relates to the right rectangular barrier located in the interval
$[a_2,b_2]$.

Calculations yield (see Section \ref{back})
\begin{eqnarray*}
\fl \tau^{(1)}_{tr}=\tau^{(2)}_{tr}=\frac{m}{4\hbar k\kappa^3}\left[2\kappa
d(\kappa^2-k^2)+\kappa_0^2\sinh(2\kappa d)\right],\\
\fl \tau^{gap}_{tr}=\frac{m}{\hbar k^2T}\left[k
L(1+R)+4\eta\sqrt{R}\sin\left(\frac{kL}{2}\right)\sin\left(J+\frac{kL}{2}\right)\right],\\
\fl \tau^{(1)}_{ref}=\frac{mT_{two}}{2\hbar k\kappa^3}\Big\{2\kappa
d\left[\kappa^2-k^2-\kappa_0^2\cos(kL)\right]+4k\kappa\sin(kL)\sinh^2(\kappa d)\\
\fl +\left[\kappa_0^2-(\kappa^2-k^2)\cos(kL)\right]\sinh(2\kappa d)
\Big\}|P|^2,\\
\fl \tau^{gap}_{ref}(k)=\frac{mT_{two}}{\hbar k^2}\Big[kL-\sin(kL)\Big]|P|^2;
\end{eqnarray*}
here $|P|^2=[1+R-2\eta\sqrt{R}\sin(J+kL)]/T$ (see Exp. (\ref{6})).

Note that $\tau^{dwell}_{tr}(k)\neq\tau^{dwell}_{ref}(k)$ while $\tau^{as}_{tr}(k)=\tau^{as}_{ref}(k)$. Another feature is that
$\tau^{(2)}_{tr}=\tau^{(1)}_{tr}\equiv\tau^{bar}_{tr}$ (see (\ref{108})). If $\tau^{left}_{tr}$ and $\tau^{right}_{tr}$ denote the transmission
dwell times for the intervals $[a_1,x_c]$ and $[x_c,b_2]$, respectively, then
\begin{eqnarray} \label{401}
\fl \tau^{left}_{tr}=\tau^{right}_{tr}=\tau^{bar}_{tr}+\tau^{gap}_{tr}/2= \tau^{dwell}_{tr}/2.
\end{eqnarray}
That is, this time scale obeys the natural physical requirement: for any barrier structure possessing the mirror symmetry, the transmission time
to describe the stationary scattering process must be the same for its two reflection symmetric parts.

For comparison we present also the Buttiker dwell time $\tau_{dwell}=\frac{m}{\hbar k}\int_{a_1}^{b_2}\left|\Psi_{tot}(x,k)\right|^2dx$. Again,
let $\tau_{dwell}=\tau_{tot}^{(1)}+\tau_{tot}^{gap}+\tau_{tot}^{(2)}$ where the contributions $\tau_{tot}^{(1)}$, $\tau_{tot}^{(2)}$ and
$\tau_{tot}^{gap}$ describe, respectively, the left and right barriers as well as the gap between them. Then
\begin{eqnarray} \label{402}
\fl \tau_{tot}^{(1)}=\frac{m}{4\hbar k\kappa^3}\bigg\{2\kappa
d\left[(\kappa^2-k^2)(1+R_{two})+2\sqrt{R_{two}}\kappa_0^2\sin(J_{two}-F_{two})\right]+\nonumber\\
\fl \left[\kappa_0^2(1+R_{two})+2\sqrt{R_{two}}(\kappa^2-k^2)\sin(J_{two}-F_{two})\right]\sinh(2\kappa
d)\nonumber\\
\fl -8k\kappa\sqrt{R_{two}}\cos(J_{two}-F_{two})\sinh^2(\kappa d)\bigg\}\nonumber\\
\fl \tau_{tot}^{gap}=\frac{mT_{two}}{\hbar k^2 T}\left[kL(1+R)+2\eta\sqrt{R}\sin(J+kL)\sin(kL)\right],\ooo
\tau_{tot}^{(2)}=\tau_{tr}^{(2)}T_{two}.
\end{eqnarray}
As is seen, this concept does not possess the property (\ref{401}).

\section{Numerical results and discussion} \label{Hart}

So, we have introduced six characteristic times: the transmission and reflection dwell times to characterize the stationary scattering process, as
well as the local and asymptotic transmission group times to characterize the OCS. For the asymptotic transmission and reflection group times of
narrow in $k$ space wave packets (that both equal to $\tau_{as}$) as well as for the transmission $\tau^{dwell}_{tr}$ and reflection
$\tau^{dwell}_{ref}$ dwell times we have obtained explicit expressions. Our next step is to compare these time scales with the Buttiker dwell time
$\tau_{dwell}$ and Wigner phase time $\tau_{ph}$; both are treated in the tunneling time literature as tunneling times and both predict the
Hartman effect.

\begin{figure}[t]
\begin{center}
\includegraphics[width=6.5cm,angle=0]{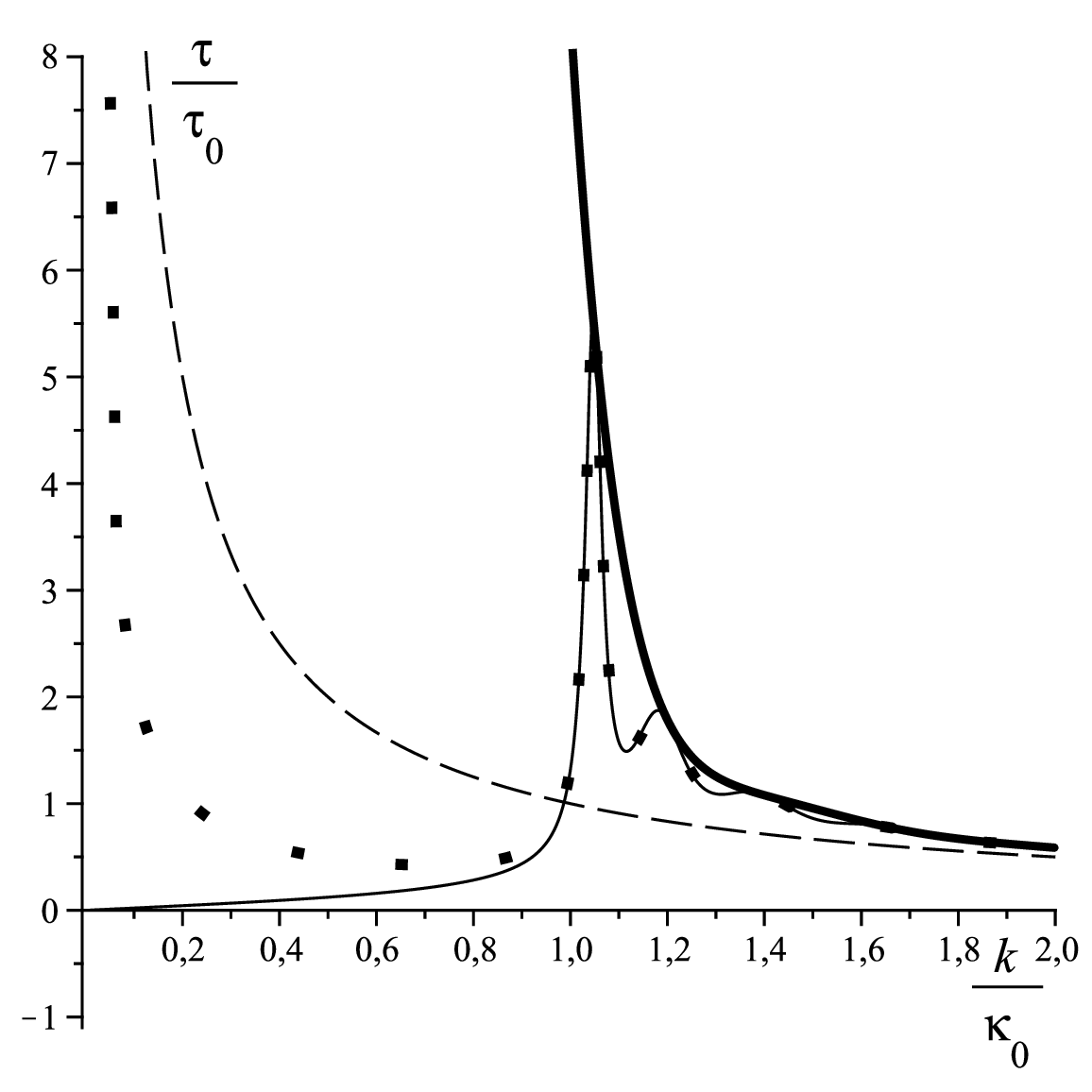}
\end{center}
\caption{ $\tau^{dwell}_{tr}$ (bold full line), $\tau_{dwell}$ (full line), $\tau_{ph}$ (dots) and $\tau_{free}=mD/\hbar k$ (broken line) as
functions of $k$ for a system with $2\kappa_0 d =3\pi $ and $L=0$; $\tau_0=2md/\hbar\kappa_0$ (see also fig.~3 in \cite{But}).} \label{fig.1}
\end{figure}

\begin{figure}[h]
\begin{center}
\includegraphics[width=6.5cm,angle=0]{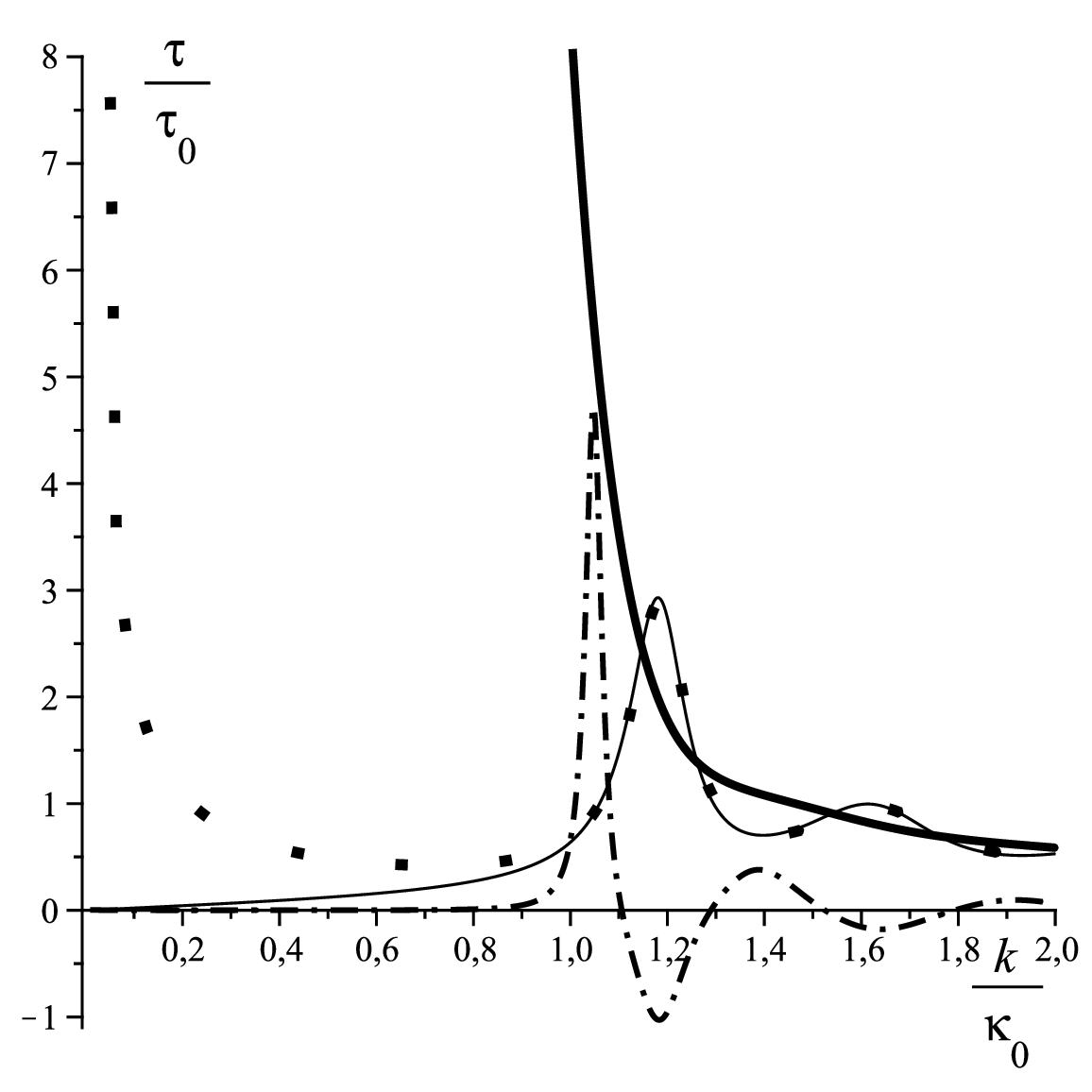}
\end{center}
\caption{$\tau^{dwell}_{tr}$ (bold full line), $\tau_{dep}$ (dash-dot), $\tau^{dwell}_{ref}$ (full line) and $\tau_{as}$ (dots) as functions of
$k$; all parameters of a system as in fig.~\ref{fig.1}} \label{fig.2}
\end{figure}

\begin{figure}[t]
\begin{center}
\includegraphics[width=6.5cm,angle=0]{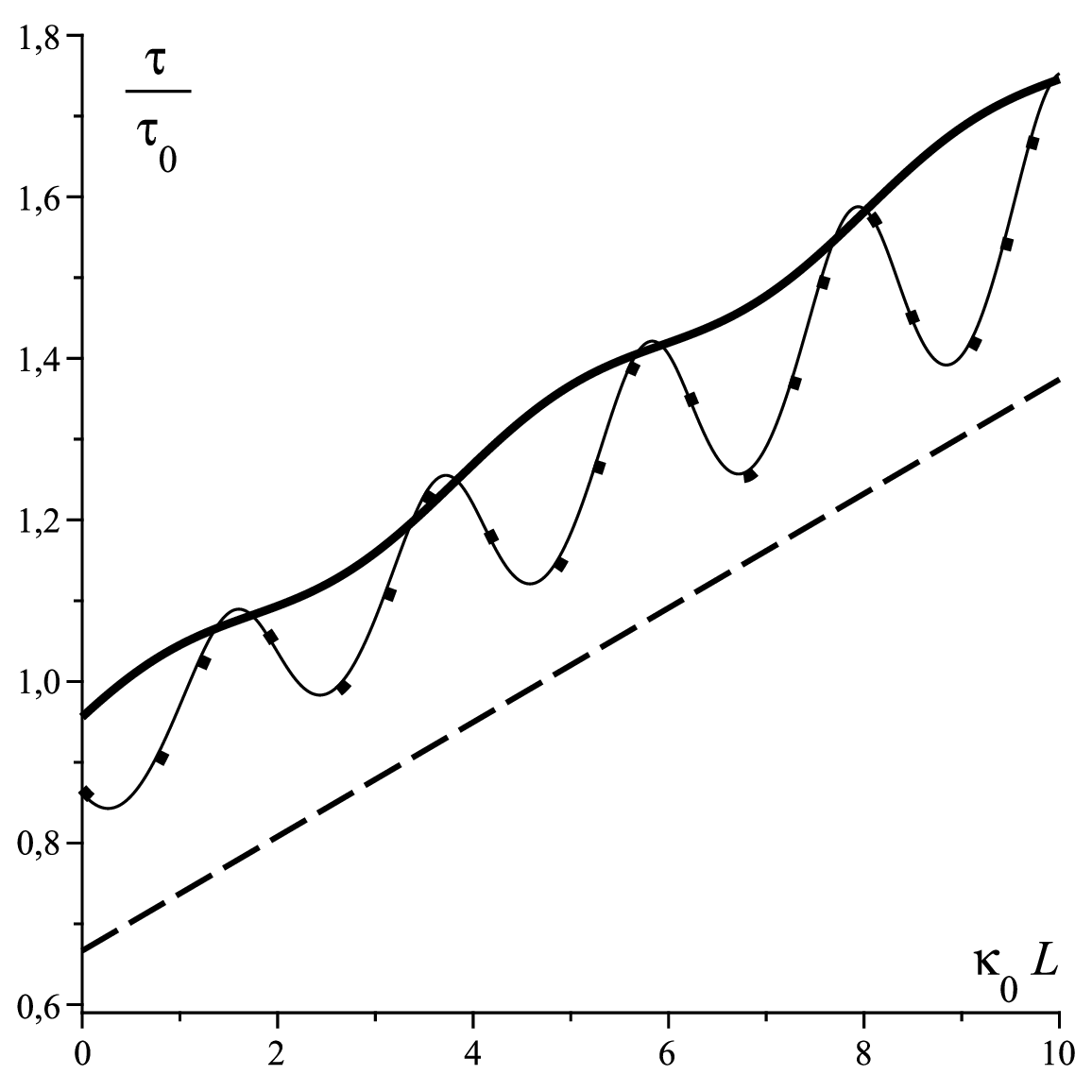}
\end{center}
\caption{$\tau^{dwell}_{tr}$ (bold full line), $\tau^{dwell}$ (full line), $\tau_{as}$ (dots) and $\tau_{free}$ (broken line) as functions of $L$
for $2\kappa_0 d =3\pi $ and $k=1.5 \kappa_0$.} \label{fig.3}
\end{figure}

\begin{figure}[h]
\begin{center}
\includegraphics[width=6.5cm,angle=0]{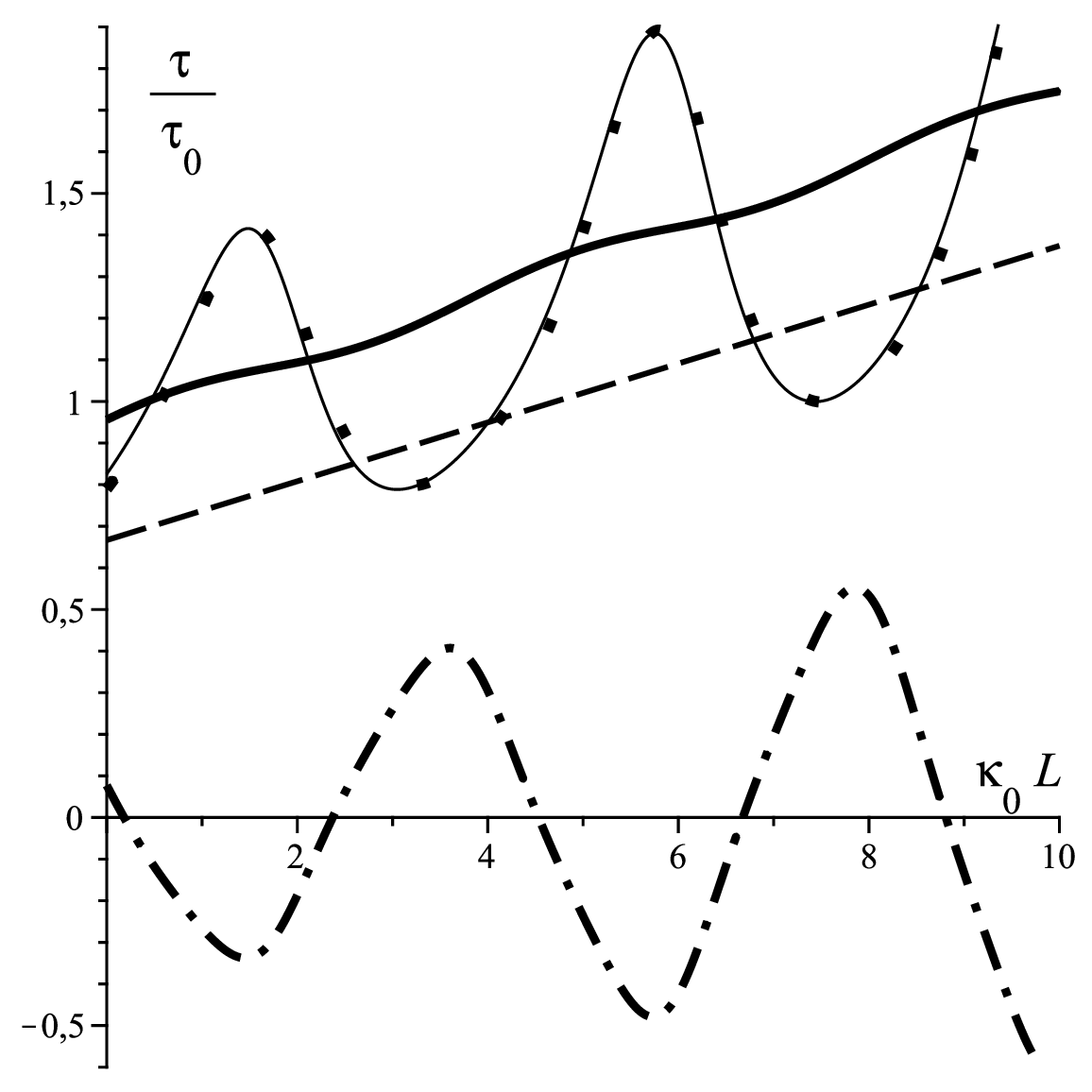}
\end{center}
\caption{$\tau^{dwell}_{tr}$ (bold full line), $\tau^{dwell}_{ref}$ (full line), $\tau_{as}$ (dots), $\tau_{dep}$ (dash-dot) and $\tau_{free}$
(broken line) as functions of $k$ for the same system and particle energy as in fig.~\ref{fig.3}} \label{fig.4}
\end{figure}

\begin{figure}[t]
\begin{center}
\includegraphics[width=6.5cm,angle=0]{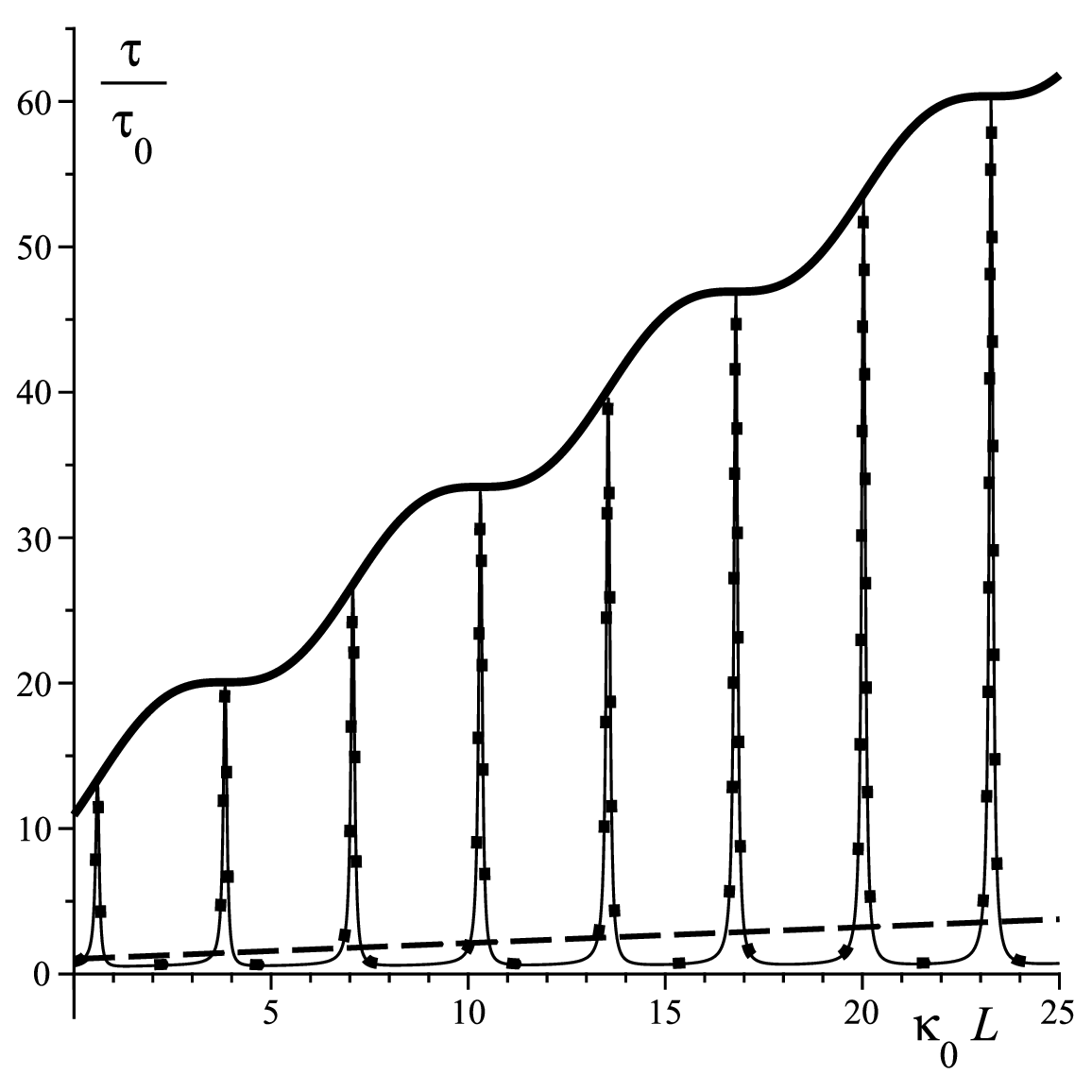}
\end{center}
\caption{$\tau^{dwell}_{tr}$ (bold full line), $\tau^{dwell}$ (full line), $\tau_{as}$ (dots) and $\tau_{free}$ (broken line) as functions of $L$
for $2\kappa_0 d =3\pi $ and $k=0.97 \kappa_0$.} \label{fig.5}
\end{figure}

\begin{figure}[h]
\begin{center}
\includegraphics[width=6.5cm,angle=0]{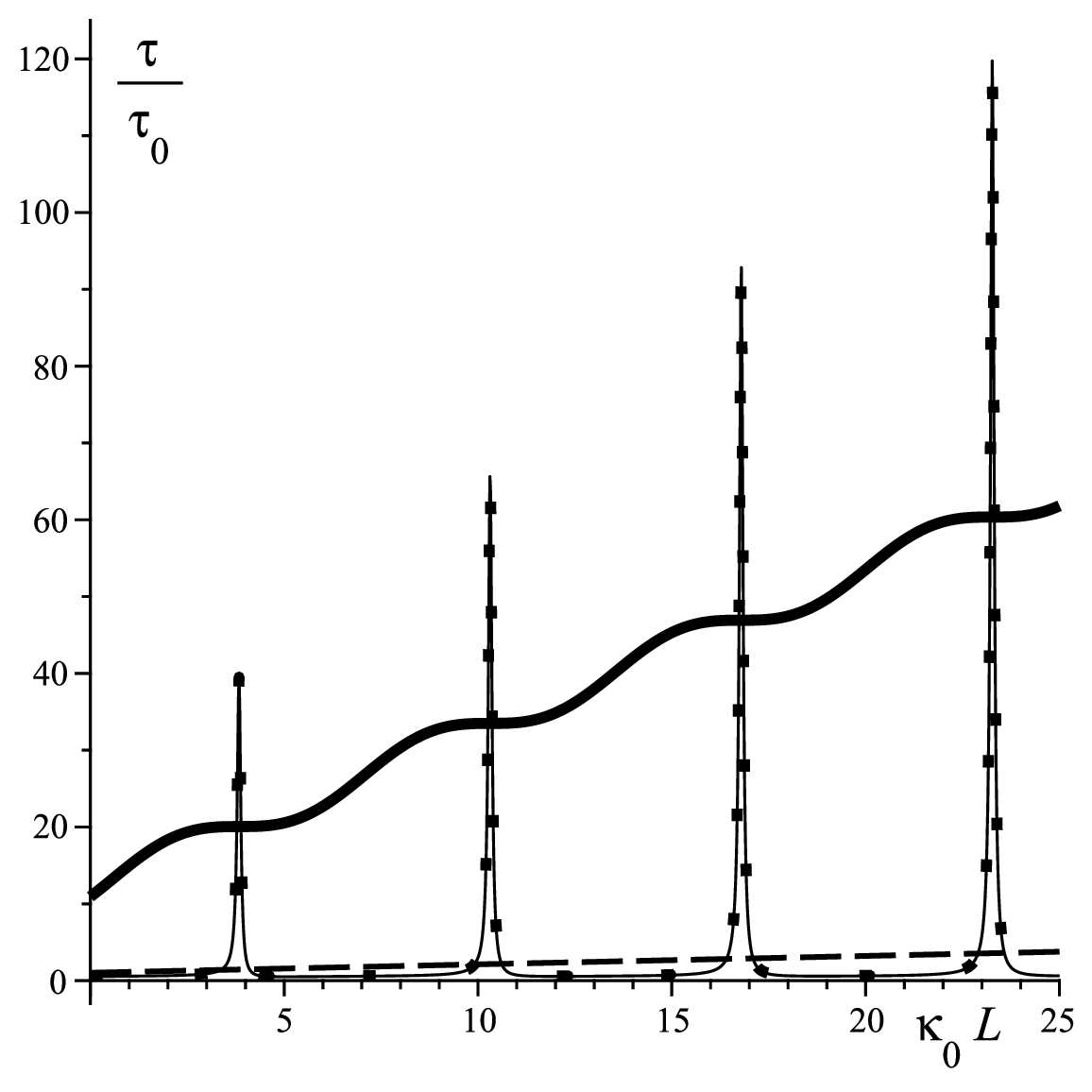}
\end{center}
\caption{$\tau^{dwell}_{tr}$ (bold full line), $\tau^{dwell}_{ref}$ (full line), $\tau_{as}$ (dots) and $\tau_{free}$ (broken line) as functions
of $k$ for the same system and particle energy as in fig.~\ref{fig.5}} \label{fig.6}
\end{figure}

As is known, $\tau_{ph}$ diverges and $\tau_{dwell}$ diminishes in the low energy domain, but both approach each other in the high energy domain
(see, e.g., fig.~3 in \cite{But}). In our approach, the same connection exists between $\tau_{as}$ and $\tau^{dwell}_{ref}$ (see
figs.~\ref{fig.1}-\ref{fig.6}). In all these figures, the quantity $\tau^{dwell}_{tr}/\tau_0$ is presented as a 'reference' one. Unlike the
conventional time scales $\tau_{ph}$ and $\tau_{dwell}$, as well as our $\tau_{as}$, the transmission dwell time $\tau^{dwell}_{tr}$ never leads
to nonphysical, anomalously short tunneling times.

As is seen from figs.~\ref{fig.1} and \ref{fig.2}, all the analyzed time scales approach the free-passage time $\tau_{free}=m L/\hbar k$ in the
high energy domain. However, in the low energy domain, $\tau^{dwell}_{tr}\gg \tau_{as} \approx\tau_{ph}\gg \tau^{dwell}_{ref}\approx\tau_{dwell}$.
Here the departure time $\tau_{dep}$ diminishes, as in the high energy domain, and hence our approach justifies the concept of the Wigner
tunneling time for particles with sufficiently high and low energies. This takes place also at the points to lie between resonances on the whole
energy axis. As regards the very resonance points, here $|\tau_{ph}-\tau_{as}|=|\tau_{dep}|$ is maximal (see fig.~\ref{fig.2} and
fig.~\ref{fig.4}).

Note that the function $\tau_{dwell}(k)$ intersects the one $\tau^{dwell}_{tr}(k)$ at all resonance points. Like the phase time $\tau_{as}(k)$ it
takes maximal values in the vicinities of resonance points. It is interesting that $\tau_{as}(k)$ and $\tau^{dwell}_{ref}(k)$ do this only at the
resonance points with even numbers (for example, these functions have no maximum at the lowest energy resonance). The CMs of the wave packets
$\psi_{tr}(x,t)$ and $\psi_{ref}(x,t)$, peaked on the energy scale at the resonances with even numbers, start earlier ($\tau_{dep}(k)<0$) than
that of the total wave packet $\Psi_{tot}(x,t)$. At the resonance points with odd numbers we meet an opposite situation. Moreover, at such
resonances, the local maxima of the function $\tau_{ph}(k)$ transform into the local minima of $\tau_{as}(k)=\tau_{ph}(k)-t_{tr}^{dep}(k)$.


When $E>V_0$, both tunneling and reflection times increase as $L\to\infty$ (see figs.~\ref{fig.3} and fig.~\ref{fig.4}). However, in the tunneling
regime, only the transmission dwell time $\tau^{dwell}_{tr}(k)$ monotonously increases in this case (see figs.~\ref{fig.5} and fig.~\ref{fig.6}).
Other four time scales, in between the resonance points, saturate in this case. Moreover, $\tau^{dwell}_{ref}(k)$ and $\tau_{as}(k)$ do this also
at the resonance points with odd numbers.


\begin{figure}[h]
\begin{center}
\includegraphics[width=8.0cm]{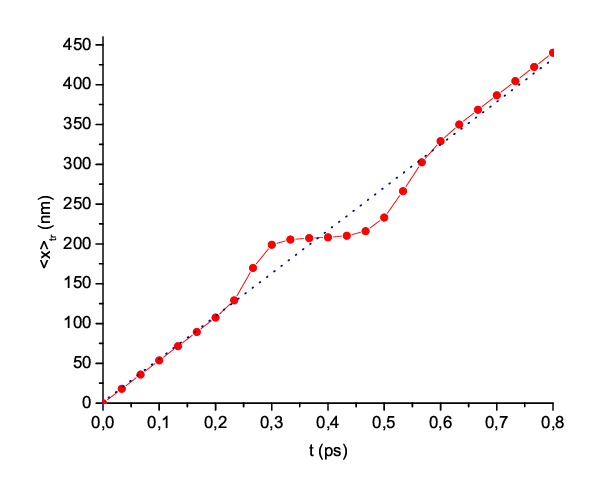}
\end{center}
\caption{The CM's positions for $\psi_{tr}(x,t)$ (circles) and for the corresponding RWP (dashed line) as
functions of time $t$.} \label{fig:fig5a1}
\end{figure}

So, in the opaque limit the transmission dwell time $\tau^{dwell}_{tr}$ is much larger than the asymptotic transmission group time
$\tau^{as}_{tr}$ which like $\tau_{ph}$ and $\tau_{dwell}$ saturates in this case. However, these two facts do not at all mean that our approach
leads to mutually contradictory tunneling times, with one of them violating special relativity. In order to understand this paradoxical situation
let us analyse the function ${\bar{x}}_{tr}(t)$ to describe scattering the Gaussian wave packet (\ref{200}) on the rectangular potential barrier
(i.e., $L=0$): $l_0=10nm$, $\bar{E}=(\hbar\bar{k})^2/2m=0.05eV$, $a_1=200nm$, $b_2=215nm$, $V_0=0.2eV$. In this case $\tau^{loc}_{tr}\approx
0,155ps$, $\tau^{as}_{tr}\approx 0,01ps$, $\tau_{free}\approx 0,025ps$ (see fig. \ref{fig:fig5a1}).

This figure shows explicitly the qualitative difference between the local $\tau_{tr}^{loc}$ and asymptotic $\tau_{tr}^{as}$ transmission group
times. While the former gives the time spent by the CM of this packet in the barrier region, the latter describes the influence of the barrier on
the CM in the course of the whole scattering process. Consequently, the quantity $\tau^{as}_{tr}-\tau_{free}$ is the time delay acquired by the CM
in the course of the whole scattering process; $\tau_{free}=mD/\hbar k$. It describes the relative motion of the CMs of the {\it transmitted} wave
packet and the corresponding freely moving reference wave packet (RWP) whose departure time is $\tau_{dep}$ which is approximately with that of
the total wave packet $\Psi_{tot}(x,t)$ when the barrier is opaque.

Thus, the influence of the opaque rectangular barrier on the transmitted wave packet has a complicated character. The {\it local} transmission
group time says that the barrier retards the motion of the CM when it enters the barrier region, while the {\it asymptotic} transmission group
time tells us that the total influence of the opaque barrier on the transmitted wave packet has accelerating character: at the final stage of a 1D
completed scattering this packet moves ahead the RWP.

Note, for any finite value of $l_0$, the velocity of the {\it CM} of the wave packet $\psi_{tr}(x,t)$ can be associated with the average velocity
of transmitted {\it particles} only at the initial and final stages of scattering. However, when the value of $l_0$ is large enough (the packet is
narrow in $k$ space) this takes place also at the very stage of scattering, when the CM of this packet moves inside the barrier region $[a_1,b_2]$
and its leading and trailing fronts move far beyond this region. At this stage, only the main harmonic $\bar{k}$ determines the input and output
probability flows at the point $x_c$. As a result, these flows balance each other, and hence the norm ${\textbf{T}}$ is constant at this stage. In
this case the local transmission group time, like the transmission dwell time, allows us to reveal the average velocity of tunneling particles.
And both these time scales show the effect of retardation of tunneling particles in the barrier region $[a_1,b_2]$, in the opaque limit.

Another situation arises when the leading or trailing front of the wave packet $\psi_{tr}(x,t)$ crosses the point $x_c$. Namely, when its leading
front crosses this point this point acts as a 'source' of particles, resulting in the acceleration of the CM, located at this stage to the left of
the structure. When its trailing front passes this point the latter acts as a 'sink', again leading to the acceleration of the CM, which is
located now to the right of the structure (see Fig~\ref{fig:fig5a1}). It is this acceleration effect that leads, in the opaque limit, to the
saturation of the transmission group time and superluminal tunneling velocities. However, this acceleration does not at all mean that {\it
particles} accelerate at these stages.

The main feature of transmission is that, like reflection, it is only a part of the OCS. Thus, it cannot be {\it directly} observed because
transmission is inseparable from reflection. And, at first glance, this fully concerns the reflection subprocess. However, this is not. It is not
occasional that the norm of $\psi_{ref}(x,t)$ is constant at all stages of scattering (see Section \ref{alt}). That is, this subprocesses is
unitary as the whole process OCS. And, as the OCS, it can be observed directly. Namely, it can be directly observed in the region $x<x_c$ in the
case of a bilateral scattering described by the wave function $\Psi_{ref}(x,t)$: $\Psi_{ref}(x,t)\equiv \psi_{ref}(x,t)$ for $x<x_c$ and
$\Psi_{ref}(x-x_c,t)=-\psi_{ref}(x_c-x,t)$ for $x>x_c$.

Thus, in principle, one can read the equality (\ref{203}) as $\psi_{tr}(x,t)=\Psi_{tot}(x,t)-\psi_{ref}(x,t)$ and consider transmission as a
result of superposition of the whole process of the OCS and its reflection subprocess, both being directly observable. This means that the above
superluminal motion of the CM of $\psi_{tr}(x,t)$ is an (irremovable) interference effect. And, what is important is that this effect takes place
even when the transmission group velocity is subluminal. Thus, the concept of the asymptotic transmission group time $\tau^{as}_{tr}$ does not
allow one to reveal the (average) velocity of transmitted particles in the region $[a_1,b_2]$. The concept of the local transmission group time
$\tau^{loc}_{tr}$ is too a bad 'assistant' in this matter: in the case of the wave packets, narrow in $k$ space, the CM's position in this region
cannot be defined with a proper accuracy; in the general case, the non-conservation of the number of particles at the point $x_c$ can be essential
during the whole stage of interaction of the wave packet with this point. This means that, for particles with a well defined energy, only the
concept of flow velocity that underlies the time scale $\tau^{dwell}_{tr}$ can be used for revealing their tunneling velocity:
$\tau^{dwell}_{tr}$, as an additive quantity, is unaffected by the processes taking place at the point $x_c$.

However, of importance is once more to stress that, for transmission, neither the anomalously short asymptotic group time nor the huge dwell time
cannot be measured directly. Yes, our approach confirms that superluminal group tunneling velocities, observed in the tunneling time experiments,
indeed relate to the inherent properties of tunneling. But these measurements cannot be considered as {\it direct} ones before an experimentalist
has not proven that the reference wave packet used in the experimental timekeeping procedure to underlie his experiment can indeed be considered,
at the initial stage of scattering, as a wave packet causally connected to the transmitted one.

The well known Larmor-clock procedure \cite{But}, too, does not allow any direct measurement of the tunneling time. According to \cite{Ch6}, the
Larmor precession is not a single physical process to influence the average spin of (to-be-)transmitted particles in the region $[a_1,b_2]$. Again
the joining point $x_c$ plays crucial role: the electron spin averaged over the superposition $\psi_{tr}(x,t)=\Psi_{tot}(x,t)-\psi_{ref}(x,t)$
undergos flipping at the joining point $x_c$. As a result, the difference between the final and initial readings of the Larmor clock gives the sum
of the transmission dwell time and the additional term to describe the flipping effect. In the opaque limit the input of this effect is negative
by sign and large by absolute value. As a result, the Larmor clock show anomalously short times, though the transmission dwell time is very large
in this case.

\section{Conclusion}

We develop a new model of scattering a quantum particle on a system of two identical rectangular potential barriers and obtain explicit
expressions for the dwell and asymptotic group times to characterize its subprocesses, transmission and reflection, for a particle with a well
defined energy. According to our approach, only the transmission dwell time is associated with the time spent, on average, by transmitted
particles in the barrier region. In the opaque limit, this characteristic time increases exponentially, while the asymptotic transmission group
time saturates like the Wigner phase time. Thereby our approach does not confirm the prediction of the Hartman effect made in the existing
approaches on the basis of the dwell time, but justifies its prediction on the basis of the Wigner phase time. As was shown, this effect does not
contradict special relativity, because the transmission {\it group} velocity, because of irremovable interference effects, does not coincide with
the average velocity of transmitted {\it particles} when the wave packet to describe the transmission dynamics interacts with the two-barrier
system.

At the resonance points on the energy scale, the departure time of transmitted particles does not coincide with that of the whole ensemble of
particles. Thus, the concept of the Wigner time based on the assumption of their coincidence is invalid in this case. On the contrary, the
Buttiker dwell time gives correct values of the transmission time at such energies. In the high energy region all time scales converge to
$\tau_{free}$.

And else, since all time scales that describe the transmission subprocess admit only indirect measurements, experimental data obtained in the
tunneling-time experiments cannot be properly processed and unambiguously interpreted when the transmission dynamics at all stages of scattering
remains unknown. We hope that the presented model gives a correct solution to this problem.

\section*{Acknowledgments}

This work has been partially financed by the Programm of supporting the leading scientific schools of RF (grant No 224.2012.2).

\section*{References}

\bibliography{<your-bib-database>}



\end{document}